\documentclass[pss]{wiley2sp} % provides pss two-column style
\usepackage{amsmath,amssymb}
\begin{document}
\abovedisplayskip=4.5pt
\belowdisplayskip=4.5pt

\def\Vec#1{\mbox{\boldmath $#1$}}
\def\vS{\Vec S}
\def\vT{\Vec T}
\def\rc{{\rm c}}
\def\rl{{\rm l}}
\def\rr{{\rm r}}
\def\ud{{\uparrow \atop \downarrow}}
\def\du{{\downarrow \atop \uparrow}}
\def\uu{{\uparrow \atop \uparrow}}
\def\dd{{\downarrow \atop \downarrow}}
\def\cH{{\mathcal H}}
\def\cHeff{{\mathcal H}_{\rm eff}}
\def\ave#1{\langle#1\rangle}
\def\ket#1{|#1\rangle}

% Title of the article
\title{Anomalous behavior of the spin gap of a spin-1/2 two-leg antiferromagnetic ladder with
Ising-like rung interactions}

% Abbreviated title for the page headers
\titlerunning{Spin gap of a spin-1/2 two-leg antiferromagnetic ladder}

% Authors
\author{%
  Takashi Tonegawa\textsuperscript{\Ast,\textsf{\bfseries 1}},
  Kiyomi Okamoto\textsuperscript{\textsf{\bfseries 2}},
  Shunsuke C. Furuya\textsuperscript{\textsf{\bfseries 3}},
  T\^oru Sakai\textsuperscript{\textsf{\bfseries 4,5}}}

% Abbreviated list of authors for the page headers
\authorrunning{Tonegawa et al.}

%E-mail-address of corresponding author
\mail{e-mail
  \textsf{tone0115@vivid.ocn.ne.jp}}

% author's affiliations/addresses
\institute{%
  \textsuperscript{1}\,Professor Emeritus, Kobe University, Kobe 657-8501, Japan\\
  \textsuperscript{2}\,Department of Physics, Tokyo Institute of Technology, Tokyo 152-8551, Japan\\
  \textsuperscript{3}\,DPMC-MaNEP, University of Geneva, 24 Quai Ernest-Ansermet CH-1211 Geneva, Switzerland\\
  \textsuperscript{4}\,Graduate School of Material Science, University of Hyogo, Kamigori, Hyogo 678-1297, Japan\\
  \textsuperscript{5}\,Japan Atomic Energy Agency, SPring-8, Sayo, Hyogo 679-5148, Japan
  }

\received{XXXX, revised XXXX, accepted XXXX} % do not change, will be filled in by the publisher
\published{XXXX} % do not change, will be filled in by the publisher

% Please select about four verbal keywords for your manuscript.

\abstract{%
% This is a macro for the typesetting of two-column text in an
% abstract. It will typeset the two arguments in \abstcol{}{} as the
% left and right column inside the abstract box. At the
% columnbreak there will be always a columnbreak (\par), so both
% columns start with a new paragraph. No automatic column height
% balancing is done.
%
% If used with a \titlefigure it will silently output both
% parameters as consecutive paragraphs.
%
% The macro is defined exclusively inside the argument of \abstract{};
% if used outside it will raise an error.
%
% Usage: \abstcol{<left column>}{<right column>}
\abstcol{%
Using mainly numerical methods,
we investigate the width of the spin gap of a spin-1/2 two-leg  ladder described by
$
\cH= J_\rl \sum_{j=1}^{N/2} [ \vS_{j,a} \cdot \vS_{j+1,a} + \vS_{j,b} \cdot \vS_{j+1,b} ]
          + J_\rr \sum_{j=1}^{N/2} [\lambda (S^x_{j,a} S^x_{j,b} + S^y_{j,a} S^y_{j,b}) + S^z_{j,a} S^z_{j,b}]
$,
where $S^\alpha_{j,a(b)}$ denotes the $\alpha$-component of the spin-1/2 operator
at the $j$-th site of the $a~(b)$ chain.
We mainly focus on the $J_\rr \gg J_\rl > 0$ and $|\lambda| \ll 1$ case.
The width of the spin gap between the $M=0$ and $M=1$ subspaces
($M$ is the total magnetization)
as a function of $\lambda$
 \break
  }{%
anomalously increases near $\lambda = 0$;
for instance,
for $-0.1 \lesssim \lambda \lesssim 0.1$ when
$J_{\rm l}/J_{\rm r} = 0.1$.
The gap formation mechanism is thought to be different for the $\lambda < 0$ and $\lambda > 0$ cases.
Since, in usual cases, the width of the gap becomes zero or small 
at the point where the gap formation mechanism changes,
the above gap-increasing phenomenon in the present case is anomalous.
We explain the origin of this anomalous phenomenon by use of the degenerate perturbation theory.
We also draw the ground-state phase diagram.}
}

% The class file requires the standard graphicx Latex package. See the 'LaTeX
% standard graphics and color packages documentation' for more information at
% <http://tug.ctan.org/tex-archive/macros/latex/required/graphics/grfguide.pdf>.
%
% Accepted figure file formats depend on which LaTeX flavour is used.
% Classic LaTeX is always able to use Encapsulated Postscript (EPS);
% PDFLaTeX can't use this but accepts PDF, JPG, PNG, and GIF formats.
%
% See examples for implementing graphics in floating figure environments later in this file.
% If \titlefigure is given, it takes as its mandatory parameter the
% name (without extension) of some figure file.
%\titlefigure[height=3.1cm]{empty2w}
%\titlefigurecaption{%
%  This is the caption of the \emph{optional} abstract figure. If
%  there is no abstract figure here, the abstract text should be divided into both columns.}

\maketitle   % please do not remove

\section{Introduction}

\begin{figure}[htb]%
       \centerline{
       \scalebox{0.25}{\includegraphics{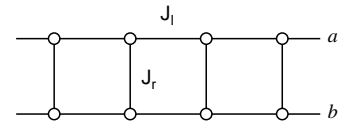}}}
\caption{Sketch of the present model.}
\label{fig:simple-ladder}
\end{figure}

Many papers on the spin ladder have been published so far \cite{works}.
Here we study the ground-state (GS) properties
of an $S=1/2$ two-leg antiferromagnetic ladder with
Ising-like rung interactions,
which has not been investigated in detail until now.
Our model is sketched in Fig.\ref{fig:simple-ladder} and expressed by
\begin{eqnarray}
    \cH
    &=& J_\rl \sum_{j=1}^{N/2} [ \vS_{j,a} \cdot \vS_{j+1,a} + \vS_{j,b} \cdot \vS_{j+1,b} ] \nonumber\\
    &&~~~   + J_\rr \sum_{j=1}^{N/2} [\lambda (S^x_{j,a} S^x_{j,b} + S^y_{j,a} S^y_{j,b}) + S^z_{j,a} S^z_{j,b}],
\end{eqnarray}
\vskip-3.5cm
\noindent
where $S^\alpha_{j,a(b)}$ denotes the $\alpha$-component ($\alpha = x,y,z$)
of the spin-1/2 operator
at the $j$-th site of the $a~(b)$ chain,
and $N$  the number of spins which is supposed to be a multiple of 4.
The parameter $J_\rl$ denotes the magnitude of the leg interaction which is isotropic,
and $J_\rr$ denotes that of the rung interaction with the Ising-like anisotropy
which is expressed by $\lambda$.
We suppose $J_\rl >0$ and $J_\rr >0$
and we restrict ourselves to
the $J_\rl \ll J_\rr$ and $|\lambda| \ll 1$ case.
Hereafter we set $J_\rr = 1$ as the unit energy,
In this paper we investigate the behavior of the spin gap
and determine the GS phase diagram on
the $\lambda-J_\rl$ plane mainly by use of the numerical methods.

\section{Anomalous behavior of the spin gap}

Let us consider the spin gap defined by $\Delta_{01} \equiv E_0(M=1,N) - E_0(M=0,N)$,
where $E_0(M,N)$ is the lowest energy of the $N$ spin system
in the subspace $M = \sum_j^{N/2} (S_{j,a}^z + S_{j,b}^z)$.
If $J_\rl=0$,
the GS is the direct product of the rung state,
where the rung state is either the singlet-dimer (SD) state
${1 \over \sqrt{2}}[(\ud) - (\du)]$
or the triplet-dimer (TD) state ${1 \over \sqrt{2}}[(\ud) + (\du)]$
depending on whether $\lambda > 0$ or $\lambda <0$.
The spin gap $\Delta_{01}$ is the energy difference between the 
SD state ($\lambda>0$) or the TD ($\lambda <0$) state
and the rung ferromagnetic state $(\uu)$,
which is equal to $(1+|\lambda|)/2$
having the minimum $1/2$ at $\lambda = 0$.
Thus, for sufficiently small leg interaction $J_\rl  \ll 1$,
it is expected that the spin gap $\Delta_{01}$ survives
and its formation mechanism is
different for the $\lambda>0$ and $\lambda<0$ cases.
The change of the gap formation mechanism is often
associated with the quantum phase transition (QPT).
In usual QPT,
we see three cases:
(a) the spin gap $\Delta_{01}$ becomes small (but finite) at the first order QPT point, 
(b) becomes zero only at the second QPT point,
(c) becomes zero in the finite range of the quantum parameter
for the Berezinskii-Kosterlitz-Thouless QPT.

%\vskip-0.3cm
\begin{figure}[htb]%
       \centerline{
       \scalebox{0.25}{\includegraphics{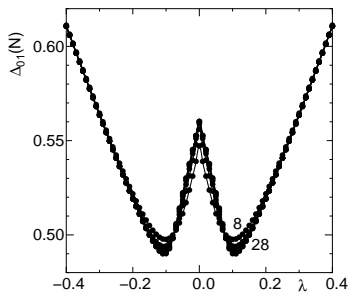}}}
\caption{Numerical results of the spin gap $\Delta_{01}$
 for $J_\rl = 0.1$ with $N = 8,12,\cdots,28$.
 The $N$ dependence of $\Delta_{01}$ is reversed between 
 the $\lambda \simeq 0.0$ case and the $\lambda \simeq 0.1$ case,
 although it is very weak.}
\label{fig:gap01}
\end{figure}
Figure \ref{fig:gap01} shows the numerical results of the spin gap $\Delta_{01}$,
calculated by the exact diagonalization (ED),
for $J_\rl = 0.1$ with $N = 8,12,\cdots,28$.
As can be easily seen,
the spin gap $\Delta_{01}$ always opens and
clearly increases in the region $-0.1 \lesssim \lambda \lesssim 0.1$,
which is quite different from the above expectations.
Thus this behavior of the spin gap $\Delta_{01}$ is very anomalous.
We note that the increase of $\Delta_{01}$ in both sides
are due to the energy difference between the SD (or TD) and the ferromagnetic state,
which is approximated by $1/2 + |\lambda|$.

\begin{figure}[htb]%
       \centerline{
       \scalebox{0.22}{\includegraphics{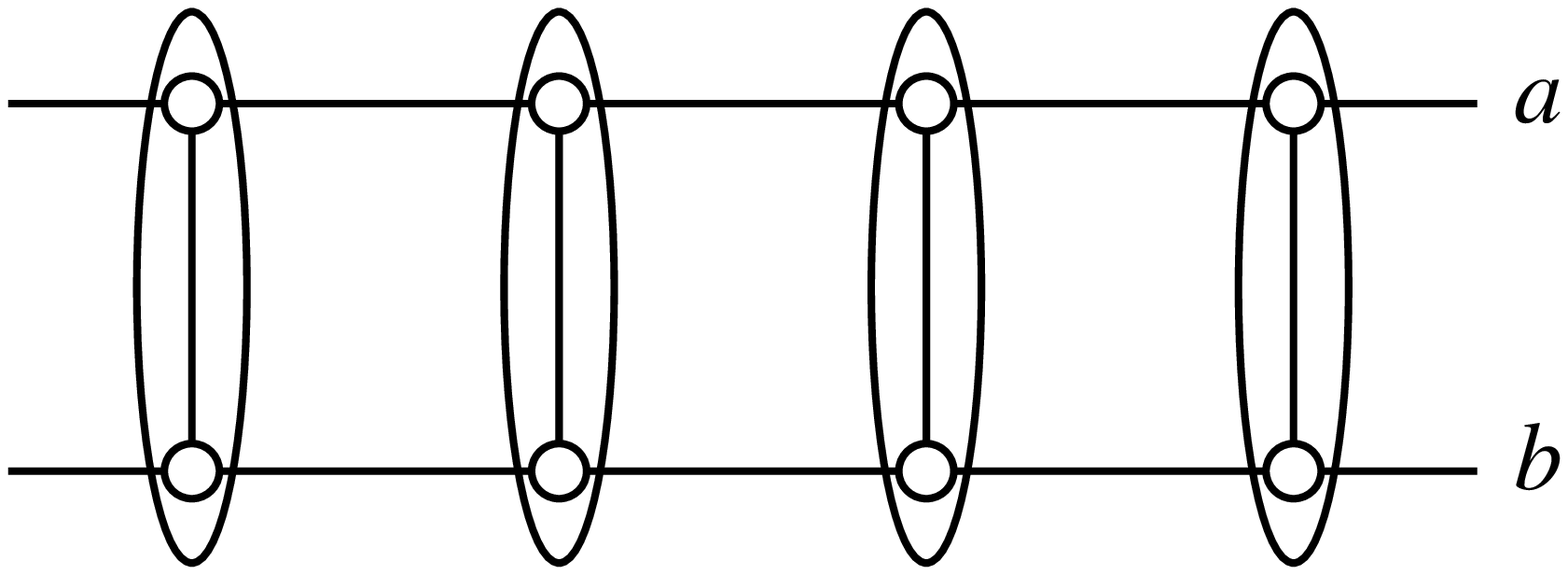}}}
       \medskip
       \centerline{
       \scalebox{0.22}{\includegraphics{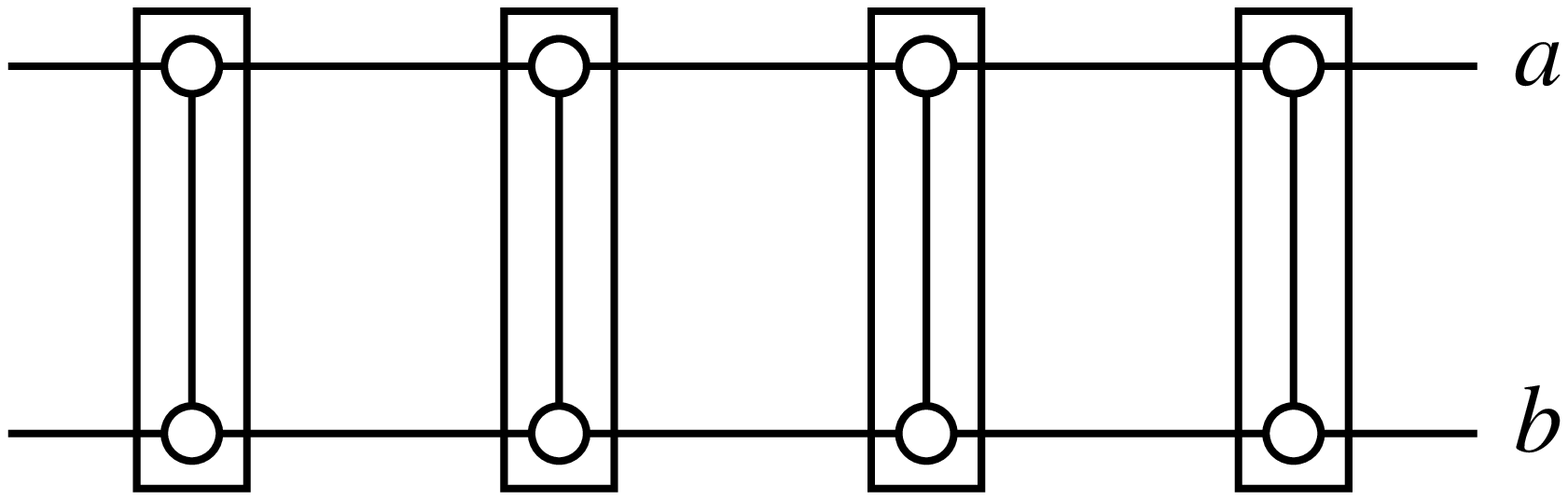}}}
       \medskip
       \centerline{
       \scalebox{0.22}{\includegraphics{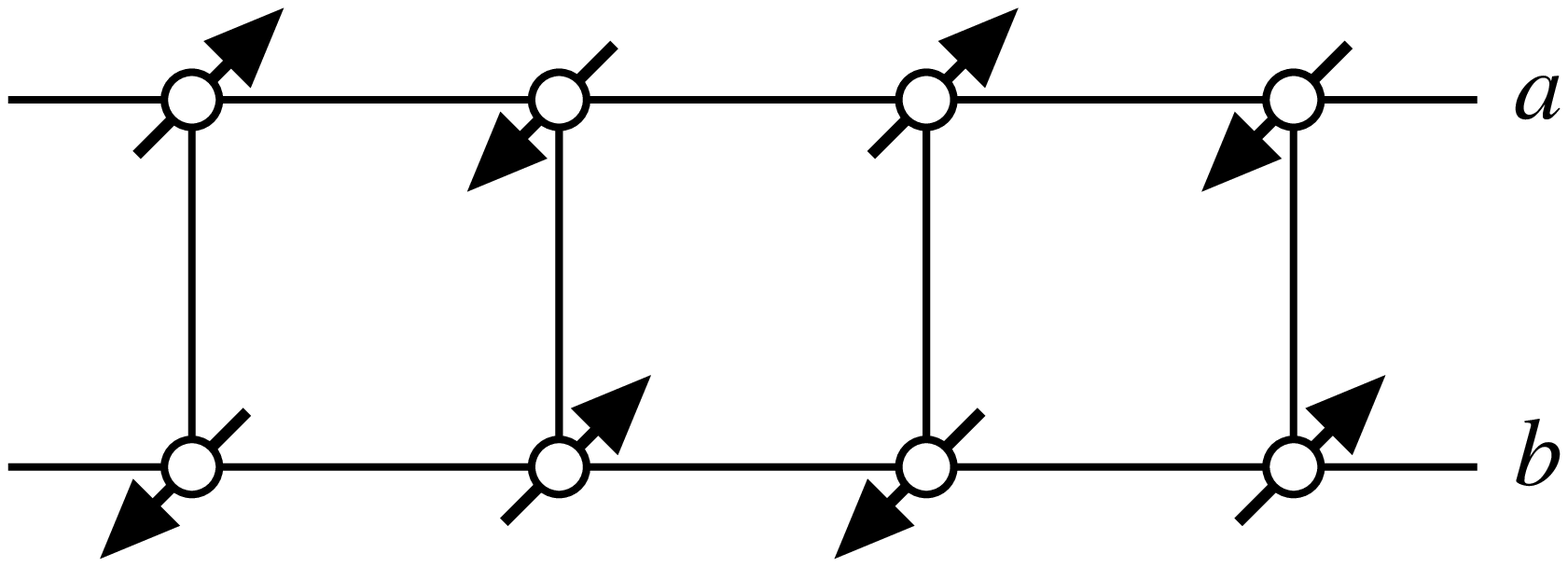}}}
\caption{Sketches of the three GSs, i.e., the SD, TD,
         and N\'eel states in the original $\vS$ picture,
         from top to bottom.
         Ellipses show the SD pairs
         and rectangles the TD pairs.}
\label{fig:gs}
\end{figure}

\begin{figure}[htb]%
       \centerline{
       \scalebox{0.3}{\includegraphics{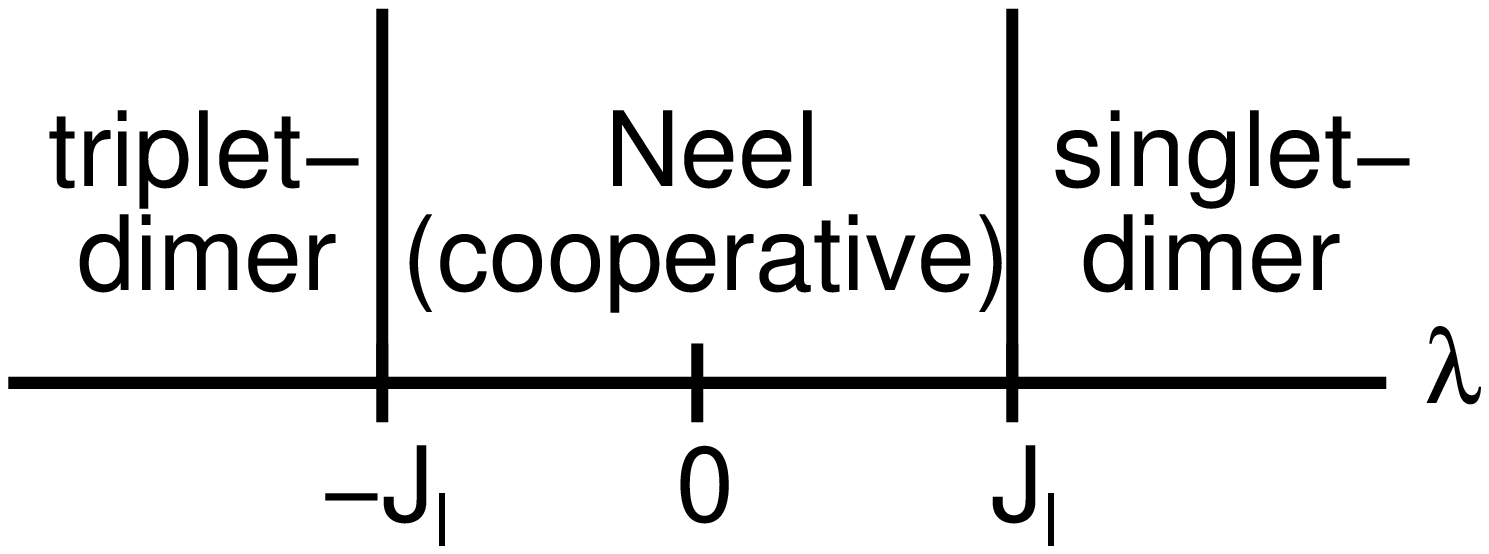}}}
\caption{GS of the present model by use of $\cHeff$.}
\label{fig:cooperative-3-mono}
\end{figure}

To explain the anomalous behavior of the spin gap $\Delta_{01}$,
we use the degenerate perturbation theory 
considering the situation $J_\rl \ll 1$ and $|\lambda| \ll 1$.
First we consider the rung state with $J_\rl = 0$,
which is nothing but the two-spin problem.
The eigenstates are the ferromagnetic states
$(\uu)$ and $(\dd)$
with the energy $1/4$,
the SD state with the energy $-(1+2\lambda)/4$,
and the TD state
with the energy $-(1-2\lambda)/4$.
Here we take the SD and TD states
into consideration,
neglecting two ferromagnetic states
with higher energies.
We introduce the pseudo-spin $\vT$ to express these two states.
Namely, the eigenstates of $T^z$ with the eigenvalues $+1/2$ and $-1/2$, 
$\ket{\Uparrow}$ and $\ket{\Downarrow}$, 
correspond to the SD state and the TD state,
respectively.
Next, by taking the effect of $J_\rl$ into account in the lowest order
we obtain the effective Hamiltonian
\begin{equation}
    \cHeff
    = 2J_\rl \sum_{j=1}^{N/2} T_j^x T_{j+1}^x
       - \lambda \sum_{j=1}^{N/2} T_j^z - {N \over 8}.
    \label{eq:heff}
\end{equation}
Since both of $\ket{\Uparrow}$ and $\ket{\Downarrow}$
have zero magnetization in the original spin picture,
%the pseudo-spin $\vT$ cannot couple with the real magnetic field and
$\cHeff$ describes only the $M=0$ subspace.
The `magnetic field term' (the second term in the rhs of eq.(\ref{eq:heff}))
comes from the energy difference between the
$\ket{\Uparrow}$ and $\ket{\Downarrow}$ states.
The effective Hamiltonian $\cHeff$ is essentially the one-dimensional
transverse field Ising model exactly solved by Pfeuty \cite{Pfeuty}.
The GS of $\cHeff$ is
either the disordered state or the N\'eel ordered state
in the $x$-direction ($x$-N\'eel state) 
according as $ |\lambda| > J_\rl$ or $ |\lambda | < J_\rl$.
The QPT between these two states is of the two-dimensional Ising type.
In the disordered phase
only the expectation value $\ave{T_j^z}$ is nonzero;
$\ave{T_j^z} > 0$ for $\lambda > 0$
and $\ave{T_j^z} < 0$ for $\lambda < 0$.
Thus, the rungs are essentially in the SD state
for $ \lambda > J_\rl$
and in the TD state for $ \lambda < - J_\rl$.
For the $x$-N\'eel phase,
the eigenstates of $T_j^x$ are
${1 \over \sqrt{2}}(\ket{\Uparrow} + \ket{\Downarrow}) = (\ud)$
and ${1 \over \sqrt{2}}(-\ket{\Uparrow} + \ket{\Downarrow}) = (\du)$
with the eigenvalues $+1/2$ and $-1/2$, respectively.
Thus the $x$-N\'eel state in the $\vT$ picture is
the $z$-N\'eel state of the original $\vS$ picture.
The N\'eel state is realized by the cooperative effect
between the SD and TD states
induced by $J_\rl$.
Three GSs are sketched in Fig.\ref{fig:gs}.
Figure \ref{fig:cooperative-3-mono} shows the GS phase diagram
based on $\cHeff$.

The physical interpretation of the above calculation is as follows.
The energy difference between the SD and TD states
per rung is $|\lambda|$.
When the perturbation energy $J_\rl$ is smaller than this energy difference,
$J_\rl < |\lambda|$,
either the SD state or the TD state becomes important for
the rung state.
On the other hand, 
when the perturbation energy $J_\rl$ is larger than this energy difference,
$J_\rl > |\lambda|$,
both of the SD and TD states are
cooperatively concerned with the formation of the rung state.
The linear combination of these two state with equal weight
brings about the rung N\'eel state.


\begin{figure}[htb]%
       \centerline{
       \scalebox{0.25}{\includegraphics{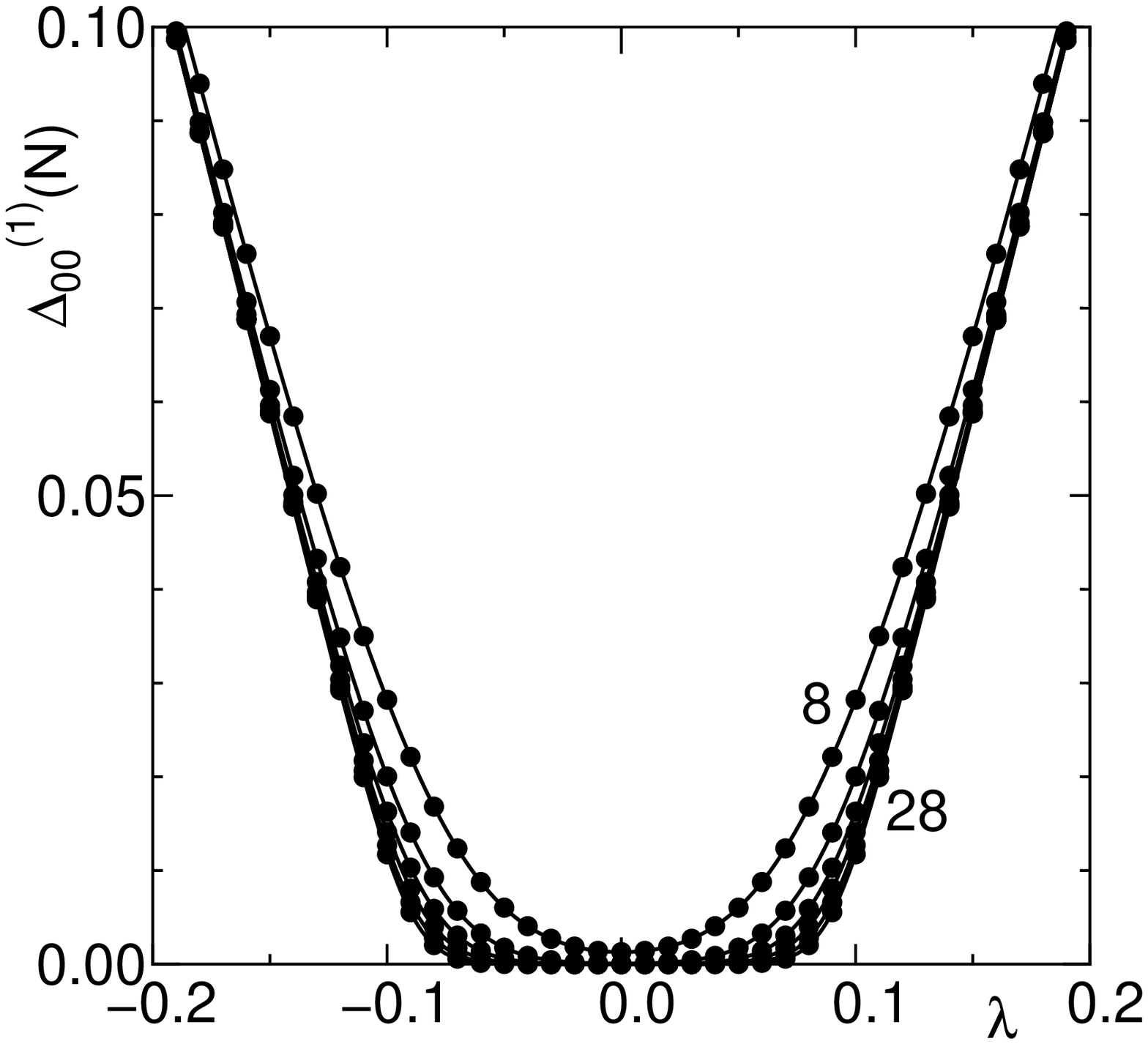}}
       \scalebox{0.25}{\includegraphics{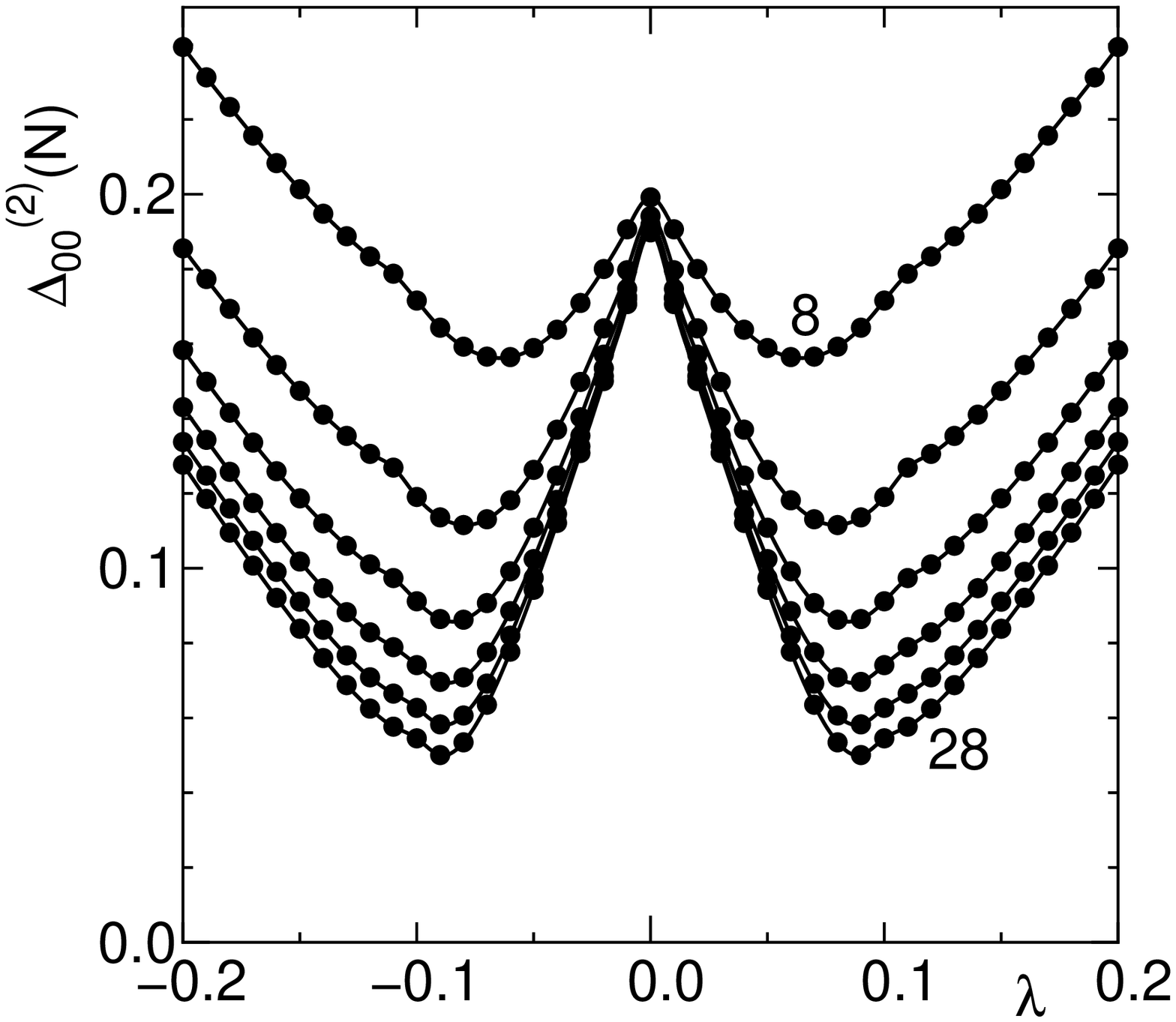}}}
\caption{Behaviors of $\Delta_{00}^{(1)}$ (left) and $\Delta_{00}^{(2)}$ (right) for the $J_\rl = 0.1$ case.
         The quantity $\Delta_{00}^{(2)}$ closes in the $N \to \infty$ limit 
         at the QTPs within the numerical accuracy.}
\label{fig:gp00p}
\end{figure}

The N\'eel state is a doubly degenerate state,
while the SD and TD states are unique and protected by symmetry.
Namely, 
if we exchange the upper and lower legs,
the wave function of the SD state changes its sign
and that of the TD state is unchanged.
For the N\'eel state,
this operation will change a wave function to the other wave function
of the doubly degenerate GS.
The GSs of finite systems are usually unique
and do not exhibit the spontaneous symmetry breaking.
For the N\'eel state,
a low-lying excited state with $M=0$ in the finite system
asymptotically degenerate to the GS as $N \to \infty$,
and the reconstruction of these two states by the linear combination
results in the doubly degenerate N\'eel state with spontaneous symmetry breaking.
Thus, in the N\'eel region,
the lowest excitation energy within the $M=0$ subspace,
$\Delta_{00}^{(1)} \equiv E_1(M=0) - E_0(M=0)$,
will rapidly decrease with the increase of the system size $N$.
Here $E_1(M=0)$ is the energy of the first excited state within
the $M=0$ subspace.
The left panel of Fig.\ref{fig:gp00p} shows the numerical results of $\Delta_{00}^{(1)}$
for $J_\rl = 0.1$ by the ED,
in which we can clearly see the above-mentioned behavior
for $-0.1 \lesssim \lambda \lesssim 0.1$.
The true gap within the $M=0$ subspace is $\Delta_{00}^{(1)}$ in the SD and TD regions,
while it is $\Delta_{00}^{(2)} \equiv E_2(M=0) - E_0(M=0)$ in the N\'eel region,
where $E_2(M=0)$ is the energy of the second excited state within
the $M=0$ subspace.
The behavior of $\Delta_{00}^{(2)}$ is shown in the right panel of Fig.5.

\begin{figure}[htb]%
       \centerline{
       \scalebox{0.25}{\includegraphics{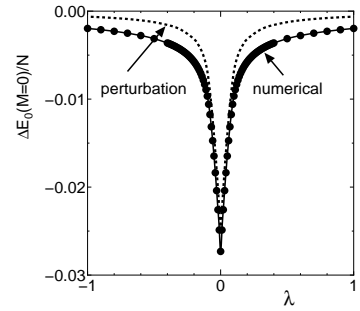}}}
\caption{Shift of the GS energy $\Delta E_0(M=0)$ for the $J_\rl = 0.1$ case. 
         Dots denote the numerical results 
         with $N=28$ 
         and broken lines the results of the perturbation theory
         eq.(\ref{eq:gs-shift}).}
\label{fig:gs-shift}
\end{figure}

Let us define the shift of the GS energy due to $J_\rl$ by
\begin{equation}
    \Delta E_0(M\!=\!0)
    \equiv E_0(M\!=\!0,J_\rl) - E_0(M\!=\!0,J_\rl\!=\!0).
\end{equation}
If we use $\cHeff$, we can obtain
\begin{eqnarray}
    &&\Delta E_0(M=0) \nonumber\\
    &&= -{N \over 2}
      \left\{ {1 \over \pi} \int_0^{\pi/2} \sqrt{J_\rl^2 + 2J_\rl \lambda \cos k + \lambda^2}\,dk
            - {\lambda \over 2}
      \right\}.      
   \label{eq:gs-shift}
\end{eqnarray}
Figure \ref{fig:gs-shift} shows the behavior of $\Delta E_0(M=0)$ for $J_\rl = 0.1$.

As can be seen in Fig. \ref{fig:gs-shift},
the decrease of the GS energy due to $J_\rl$ is much larger
in the N\'eel region than in the SD and TD regions.
The $M=1$ lowest state can essentially obtained from the GS
by changing a rung pair (N\'eel pair, SD pair
or TD pair) to the ferromagnetic pair,
although the location of the ferromagnetic pair is not fixed.
Thus,
this energy cost, 
which is essentially the spin gap $\Delta_{01}$,
is larger in the N\'eel region than
in the SD and TD regions.
This is a physical explanation for the anomalous behavior of
the spin gap $\Delta_{01}$ shown in Fig. \ref{fig:gap01}.
We note that we cannot directly calculate the spin gap $\Delta_{01}$
by use of the effective Hamiltonian $\cHeff$,
because it describes only the $M=0$ subspace.

\section{GS phase diagram}
\begin{figure}[htb]%
       \centerline{
       \scalebox{0.31}{\includegraphics{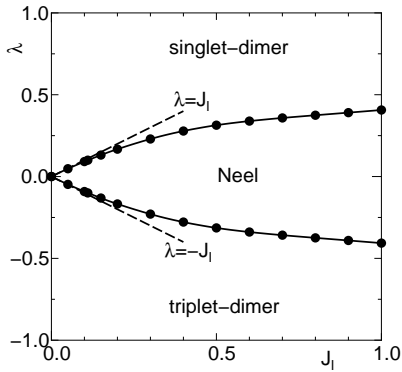}}}
\caption{GS phase diagram of the present model determined by the ED with the PRG method.
         Dots denote the numerical results and broken lines the results of the perturbation theory
         $\lambda = J_\rl$.}
\label{fig:pd1}
\end{figure}

The boundary between the N\'eel and SD (or TD) phases
can be numerically determined by use of the phenomenological renormalization group
(PRG) method \cite{prg}.
Namely, the critical value $\lambda_\rc$ for fixed $J_\rl$ can be obtained by
$\lambda_\rc = \lim_{N \to \infty} \lambda_\rc(N,N+4)$,
where $\lambda_\rc(N,N+4)$ is the numerical solution of
$N \Delta_{00}^{(1)}(N,\lambda) = (N+4) \Delta_{00}^{(1)}(N+4,\lambda)$.
Figure \ref{fig:pd1} shows the phase diagram of the present model on the $\lambda-J_\rl$ plane.
The spin gaps $\Delta_{00}^{(1)}$ and $\Delta_{00}^{(2)}$
within the $M=0$ subspace close on the phase transition lines,
whereas the spin gap $\Delta_{01}$ always opens.
Thus, on the phase transition lines, 
at sufficiently low temperatures,
the specific heat $C(T)$ shows the gapless behavior,
while the magnetic susceptibility $\chi_z(T)$ for the 
magnetic field along the $z$ direction shows a gapful behavior.
In usual cases,
both of the spin gaps within the $M=0$ subspace and that between the $M=0$ and $M=1$ subspaces
close simultaneously
on the phase transition line,
which leads to the gapless behaviors of both of $C(T)$ and $\chi_z(T)$
at low temperatures.

\section{Discussion and concluding remarks}
\begin{figure}[htb]%
       \centerline{
       \scalebox{0.25}{\includegraphics{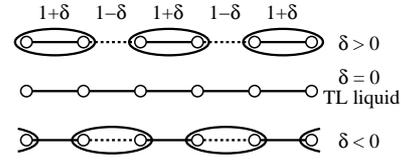}}}
\caption{GSs of the $S=1/2$ bond-alternating
Heisenberg chain $\cH_{\rm ba}$.
Ellipses denote the SD pairs.
The locations of ellipses are different for the $\delta>0$
and $\delta<0$ cases.}
\label{fig:reconstruction}
\end{figure}

Let us compare the present model with the 
$S=1/2$ bond-alternating Heisenberg chain
\begin{equation}
    \cH_{\rm ba}
    = \sum_{i=1}^N  [1+(-1)^j \delta] \vS_{j,a} \cdot \vS_{j+1,a}.
\end{equation}
where $\delta$ ($-1 \le \delta \le 1$) 
denotes the bond-alternation parameter
and $N$ the number of spins which is supposed to be even.
This bond-alternation model is a typical model
which exhibits the change of the gap formation mechanism.
When $\delta \ne 0$,
the spin gap $\Delta_{01}$ is finite even in the thermodynamic limit
as $\Delta_{01} \sim |\delta|^{2/3}/\sqrt{|\log|\delta||}$ \cite{CF,BE}.
The SD pairs exist on the $(1+\delta)$-bonds for $\delta>0$
whereas they exist on the $(1-\delta)$-bonds for $\delta<0$.
When $\delta = 0$, 
the GS is the Tomonaga-Luttinger (TL) liquid state
with the gapless excitation.
Thus the second-order QPT occurs at $\delta=0$.
This situation is sketched in Fig. \ref{fig:reconstruction}.
As can be seen from Fig. \ref{fig:reconstruction},
the location of the SD pair is different for
the $\delta>0$ and $\delta<0$ cases.
In other words,
the unit cell which is important for the gap formation
is reconstructed when $\delta$ pass through the QPT point $\delta=0$.
On the other hand,
for the  present model case,
the unit cell is always a rung
and the reconstruction of the unit cell does not occur,
as is shown in Fig. \ref{fig:gs}.
Only the change of the GS of the unit cell occurs
when $\lambda$ is swept.
This is the very important and essential point
for the quite different behavior of the spin gap $\Delta_{01}$
of the present model and the $S=1/2$ bond-alternating
Heisenberg chain model.

As stated above,
the key points to the anomalous behavior of the spin gap $\Delta_{01}$
are {\lq\lq}no reconstruction of the unit cell" and
{\lq\lq}change of the GS of the unit cell".
Thus,
in other models with these properties,
similar behavior of the spin gap $\Delta_{01}$ is expected.
In fact,
in the $S=1/2$ three-leg isosceles spin nanotube,
a similar behavior of the width of the magnetization plateau
at 1/3 of the saturation magnetization is observed \cite{3leg}.

The anomalous behavior of the spin gap $\Delta_{01}$ seems to remain even for the 
the $J_\rl \gg 1$ and $0>J_\rl \gg -1$ cases.
We have also succeeded in explaining this anomalous behavior for the former case
by use of the bosonization approach.
The details will be published elsewhere.

Finally we note that a closely related model
was investigate by Zheng-Xin Liu et al. \cite{Liu}
from a somewhat different point of view.

In conclusion,
by numerical methods,
we have observed an anomalous behavior 
of the spin gap $\Delta_{01}$ of the spin-1/2 two-leg
antiferromagnetic ladder with
Ising-like rung interactions.
Also we have succeeded in explaining the physical origin
of this anomalous behavior.

\begin{acknowledgement}
This work was
partly supported by grants-in-aid for Scientific Research (B) (no.
23340109) and Scientific Research (C) (no. 23540388), 
from the Ministry of Education, Culture, Sports, Science,
and Technology of Japan.
S.C.F. acknowledges Institute for the Solid State Physics,
the University of Tokyo, 
where a part of this work was carried out.
We thank the Supercomputer Center, the Institute for Solid State Physics, 
the University of Tokyo, 
for computational facilities.

\end{acknowledgement}

% Use the following code if you wish to generate your bibliography with BibTeX;
% replace the string "pss-demo" below with the name(s) of
% the BibTeX data base(s) you want to use.
% The resulting bibliography-output (the content of the .bbl file)
% must be pasted back into this file before submission.
% Please also include your BibTeX data base file(s) in your submission
% so that we can re-run BibTeX if necessary.
%
%\bibliographystyle{pss}
%\bibliography{pss-demo}
%
% Replace the following example bibliography with your references
% before submission:

\end{document}